\documentclass[a4paper]{article}
\usepackage[utf8]{inputenc}
\usepackage[OT4,T1]{fontenc}
\usepackage{graphicx}\graphicspath{{img/}}
\usepackage{lmodern,fancyhdr,secdot,float,amsmath,amsfonts,amsthm,amssymb}
\usepackage{booktabs,tabularx}
\usepackage[margin=2cm]{geometry}
\usepackage{multicol}
\usepackage[hidelinks,unicode]{hyperref}
\usepackage{titlesec}
\titleformat{\section}{\centering\normalsize\bf}{\thesection.}{.5em}{\MakeUppercase}
\titleformat*{\subsection}{\bf\normalsize\selectfont}
\titleformat*{\subsubsection}{\bf\normalsize\selectfont}

\newcommand{\titleEN}[1]{\normalsize #1}

\newcommand{\keywordsEN}[1]{\small\textbf{Keywords:} #1}

\newcommand{\abstractEN}[1]{\small\textbf{Abstract:} #1}

\begin{document}\thispagestyle{empty}\pagestyle{fancy}

\begin{minipage}[t]{1.0\textwidth}\vspace{0pt}%
\centering
\textit{Accepted for Konferencja Radiokomunikacji i Teleinformatyki KRiT-2023, Krakow 2023 (author's version)}
\end{minipage}

\vspace{0.1cm}

\begin{center}

\titleEN{VEHICLE DETECTION IN 6G SYSTEMS WITH OTFS MODULATION}\medskip

Pavel Karpovich $^{1,2}$;
Tomasz P. Zielinski$^{2}$;

\medskip

\begin{minipage}[t]{0.4\textwidth}
\small $^{1}$ Institute of Telecommunications AGH, Krakow \href{mailto:email}{karpovic@agh.edu.pl,tzielin@agh.edu.pl}\\
\small $^{2}$ Nokia Solutions and Networks, Krakow, \href{mailto:email}{pavel.karpovich@nokia.com} \\
\end{minipage}

\medskip
\end{center}

\medskip

\begin{multicols}{2}
\noindent


\medskip

\noindent

\abstractEN{
The recently introduced orthogonal time frequency space modulation (OTFSM) is more robust to large narrow-band Doppler frequency shift than the orthogonal frequency division multiplexing (OFDM), used in the 5G standard. In this paper it is shown how the telecommunication OTFSM-based signal with random padding can be used with success in the 6G standard for detection of high-speed vehicles. Two approaches for detecting targets during the random padded OTFS based transmission are compared in the paper.}

\medskip


\noindent
\keywordsEN{5G, 6G, OFDM, OTFSM, radar.}

\section{Introduction}

In last few years, the scientific community attention has been focused on the discussion of next generation 6G communication. There are a lot of publications about what applications will drive the 6G network and what technologies should be included in the 6G standard to satisfy their requirements \cite{6G_harsh} \cite{6G_vision}. Among large number of proposals, there are some that are most common, such as a terahertz wave and an integrated sensing and communication (ISAC) \cite{isac1} \cite{isac2}. This paper addresses a problem of adding a radar functionality to the communication systems of the future which will use higher frequency carriers and support high-mobility users.  

The usage of terahertz band is challenging. Even relatively slow objects could generate very high Doppler frequency shifts. The strong Doppler effect limits the usage of the orthogonal frequency division multiplexing (OFDM) waveform which is at present de-facto a standard waveform in telecommunication systems (e.g. DVB-T2, Wi-Fi, LTE, 5G \cite{{ofdm_numerology}}). The OFDM is based on assumptions that linear convolution of the signal and the channel impulse response can be replaced by circular convolution, and that the channel impulse response is time-invariant or almost time-invariant. This allows to do a very fast and simple channel impulse response estimation. In case of the strong Doppler environment the assumption about constant channel impulse response is no longer valid since any channel coefficient can rotate in complex plane all the time due to the Doppler effect. Using OFDM in such conditions leads to errors in channel estimation and equalization, and eventually to inter-carrier-interference (ICI) and subsequently errors in bit detection.

Increasing sub-carrier spacing (SCS) in OFDM helps to deal with the strong Doppler frequency shift. However, this operation will increase also the OFDM cyclic prefix overhead and reduce transmission efficiency \cite{ofdm_numerology}. In order to eliminate the mentioned above disadvantage of the OFDM, the orthogonal time frequency and space (OTFS) modulation was recently introduced in \cite{otfs1}. Due to its unique features it is seriously treated as one of possible 6G waveforms \cite{otfs2}.

In this article simulation results for an ISAC system using the OTFS waveform are shown. We will start with the OTFS waveform description, present the delay-Doppler domain used in OTFS and discuss different pilot configurations exploited in it. Next, we will introduce the ISAC system using the OTFS waveform. Finally, in experimental part, we will show results from simulation of a radar part of the discussed RP-OTFS-based ISAC system.

In work \cite{my_rp1} results from simulation of the communication part of the RP-OTFS transmission system were presented while this paper addresses simulation of the radar part of the system only. Practical verification of the general RP-OTFS based transmission and sensing concept was already presented in \cite{my_rp2}.

\section {Orthogonal time frequency and space}

\begin{figure}[H]
\centering
\includegraphics[scale=0.5]{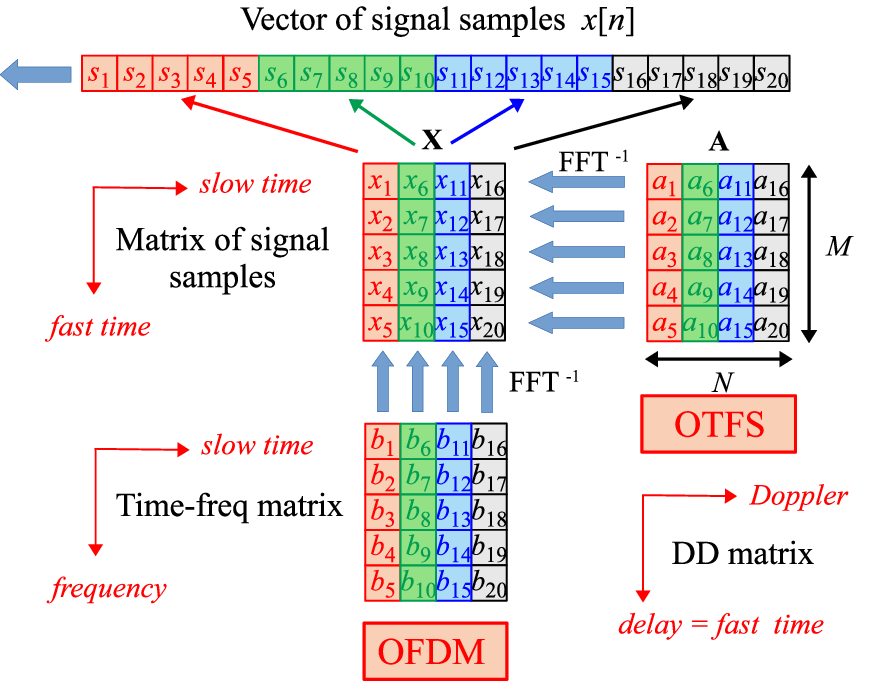}
\caption{The OTFS concept}
\label{fig_otfs}
\end{figure}

The concept of the OTFS is shown in the figure \ref{fig_otfs} \cite{my_rp1} \cite{my_rp2}. In comparison to OFDM, the OTFS is a two-dimensional modulation technique. In case of OTFS the modulation process looks as follows. At the beginning modulated IQ/QAM symbols are put into elements of the matrix~\textbf{A} in figure~\ref{fig_otfs}, i.e. on the grid in a delay-Doppler (DD) domain. Then, the inverse Zak transform (inverse Fourier transform over the Doppler axis) \cite{zak} is used to transform (demodulate) data from the DD to a fast time - slow time (TT) domain. Finally, the obtained samples are reshaped from a matrix into a vector. The DD grid usage for data modulation makes the OTFS waveform attractive for ISAC since it is ``native'' domain for radars.

\subsection {The delay-Doppler grid}

Names of the DD grid directions reflect their physical sense. The delay direction (the first D in DD) consists of adjacent samples from time domain, the $\Delta t$ between samples is small. This direction is suitable for detecting small time changes in a observed signal. For example, in multi-path propagation environment, the difference between paths is not very big and can be estimated in the delay direction of the DD grid. But the delay direction is not suitable for observation of long time processes like the Doppler effect because Doppler frequencies are usually very small and require more time for estimation. In the Doppler direction (the second D of DD) only every $M$th sample from time domain is used and this allows to estimate long time signal changes using FFTs with small sizes.  

Parameters of the DD grid should be chosen taking into account that the sent OTFS modulated waveform will be used, both, for digital data transmission and moving vehicles detection. As sources of multi-path reflections could be treated as ``radar targets'', the OTFS DD grid  should fulfill, both, telecommunication and radar requirements. The DD grid has two parameters: $M$ --- the number of samples in the delay (fast time) direction, and $N$ --- the number of samples in the Doppler  direction. Looking at figure~\ref{fig_otfs}, we can say that in Doppler direction a signal is practically decimated by $M$. Hence taking into account the Nyquist theorem, the maximum Doppler offset that can be estimated using such DD grid is $f_{d max} = \pm{f_s \over 2 M}$, where $f_s$ is a sampling rate. Resolution in the Doppler direction depends on $N$: increasing $N$ and keeping $f_s$ and $M$ constant will increase the FFT length and the Doppler resolution. The resolution in delay direction depends only on  $f_s$. Choosing $f_s$, $M$, $N$ and carrier frequency $f_c$ one can optimize the OTFS-based radar and digital transmission.

For example, lets choose the DD grid parameters for radar detection of many moving cars (reflections from stationary objects are not interesting for us). For maximum car speed of 60 m/s, carrier frequency $f_c = 52.6$~GHz (the maximum carrier frequency for 5G FR2), sampling ratio $f_s = 50$~MHz and maximum $M = 1190$, we can assume that the maximum Doppler frequency shift is equal to 21~kHz. Then, by fixing $M=1024$ and changing $N$, one can get different resolution of velocity estimation, changing from 9 m/s (for $N=8$) to 0.1 m/s ($N=512$), where values of $N=8$ and $N=512$ are exemplary ones.

\subsection {Pilots configurations}

As OFDM, the OTFS uses pilots for estimation of a channel impulse response (CIR). Their configurations are different. Here we will discuss two types of pilot placement strategies, shown in figure~\ref{fig_zprp_otfs}: a zero-padded one (ZP-OTFS) and a random-padded one (RP-OTFS). In both configurations the DD matrix \textbf{A} is divided into two parts: the data zone and the pilots zone. Every carrier in the DD grid is assigned to the pilot or data zone only, not to both of them the same time.

\subsubsection{ZP-OTFS}

\begin{figure}[H]
\centering
\includegraphics[scale=2.2]{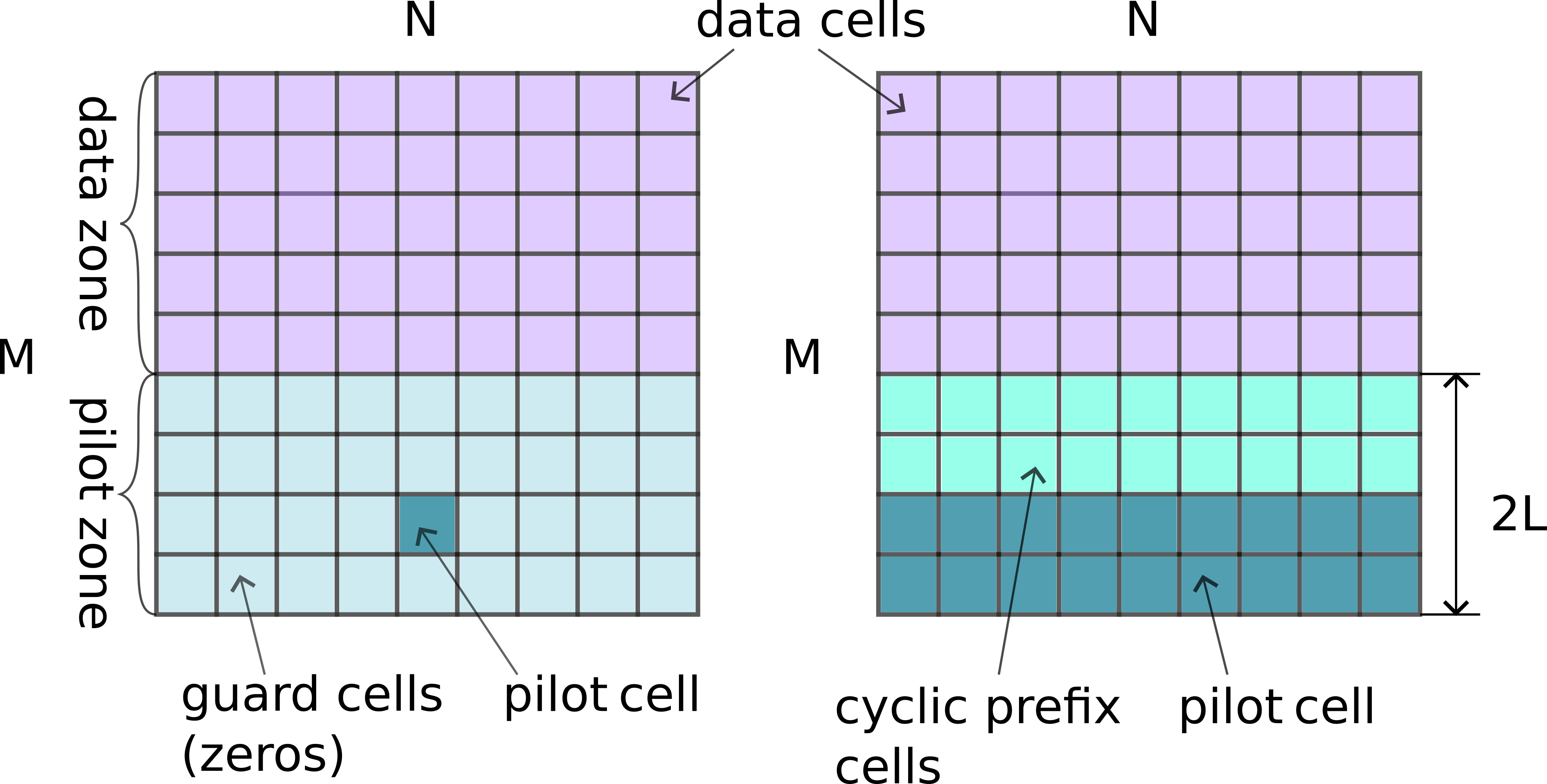}
\caption{Delay-Doppler grid of the ZP-OTFS (left) and RP-OTFS (right)}
\label{fig_zprp_otfs}
\end{figure}

In the ZP-OTFS, the pilot has a form of a rectangular zone of the DD matrix \textbf{A}, shown in figure~\ref{fig_otfs}, which is filled with zeros and have only one non-zero carrier in its center. We will call this non-zero carrier a pilot pulse.  In case of the ZP-OTFS, the length of the pilot zone in the delay direction is twice bigger than length of the channel impulse response. In the Doppler direction the pilot zone usually makes use of all cells, as shown in figure \ref{fig_zprp_otfs}. 

Due to zeros surrounding the pilot pulse, the channel estimation process becomes  very simple in the ZP-OTFS. There is also no interference between pilot and data zones as well as no ZP-OTFS symbol interference. In case of ZP-OTFS the channel impulse response is estimated by division of every cell of the received pilot zone by the known, transmitted pilot pulse (some threshold should be introduced here in order not to neglect reflection free samples of the pilot zone). The main disadvantage of the ZP-OTFS is low energy efficiency, because the pilot zone is very sparse.

\subsubsection{RP-OTFS}

The recently introduced RP-OTFS~\cite{my_rp1} \cite{my_rp2} is designed to correct deficiencies of the ZP-OTFS. Here the pilot zone is filled by short OFDM symbols, treated as pilots, with random data inside --- see figure \ref{fig_zprp_otfs}. In case of the ZP-OTFS, discussed earlier, the data zone is generated in the DD domain and transformed to fast-slow time domain by the inverse Zak transform. In turn, in case of the RP-OTFS, OFDM symbols of pilots are inserted directly into fast-slow time grid (without the inverse Zak transformation). Absence of zeros in the pilot zone increases signal-to-noise ratio (SNR) and causes that the RP-OTFS application is more efficient than the ZP-OTFS in CIR estimation what is very important for both for communication and radar.

The CIR estimation begins with conventional OFDM channel estimation with the only difference that we treat the whole OFDM symbol as a pilot. After that, when all CIR momentum estimates are found using all OFDM symbols (having transmitted and received pilots one can easily estimates CIR taps from them), we transform the matrix of CIR taps to the DD domain by the Zak transform, i.e. by performing FFT over the CIR matrix rows. Note, that in the RP-OTFS the Zak transform is performed upon CIR estimates, do not upon time samples of OFDM symbols which were used for CIR calculation.

There are two disadvantages of the RP-OTFS application. Firstly, the length of cyclic prefix (CP) of the OFDM-based pilot should be equal to the OFDM symbol length, i.e. it is long and the CP overhead reduces the achievable bit-rate. Secondly, we assume that the CIR is quasi time-invariant and, therefore, we can not use long OFDM pilots for very high frequency Doppler channels.

\section{Integrated Sensing and Communication (ISAC)}

In case of ISAC \cite{isac1}\cite{isac2}, usually, the communication processing is the same as in conventional system. In this paper we are concentrating our attention on peculiarity of RP-OTFS radar processing since efficiency of the RP-OTFS based communication sub-system has been already tested \cite{my_rp1} \cite{my_rp2}. Two approaches of target detection are analyzed:  correlation-based and pilot-based.

The first correlation-based method origins from classical radar processing in which a cross ambiguity function is used \cite{radar}: transmitted, reference signal (known, re-modulated in the receiver or acquired by special reference antenna) is shifted in time and frequency and correlated with the received, surveillance signal. The problem of the correlation base radar approach is that usually it is hard to find weak signal reflections, coming from small, moving objects, on the background of strong signal reflections caused by buildings (the radar clutter problem) \cite{my_dvbt2}.

In the second pilot-based approach of vehicle detection transmitted pilots, known in the receiver, are used to CIR estimation \cite{ofdm_base_radar}. In case of reflections coming from moving vehicles some CIR taps are complex-value numbers that oscillates in time with frequency of Doppler frequency shift caused the reflecting object movement. Here, we treat radar targets as sources of multi-path propagation. By CIR analysis we can retrieve information about signal reflections and about reflecting objects.

The pilot based ISAC system requires non distorted CIR estimates  for Doppler frequency shifts extraction. As mentioned in the introduction, high Doppler objects can not be detected by OFDM. This also limits application of pilot-based radars making use of OFDM-based pilots.

\section{Experimental part}

In experimental part we simulated a radar performance of the discussed RP-OTFS-based ISAC system. Parameters of the applied OTFS-based signal was following: size of the grid in delay and doppler direction 64x256 ($M$x$N$), length of the pilot zone 16 (meaning of $L$ is explained in fig.~\ref{fig_zprp_otfs}), modulatiotion 4-QAM, carrier frequency 4 GHz and bandwidth 20 MHz. In simulation we used different target velocities in order to test the system performance in different conditions.

\begin{figure}[H]
\centering
\includegraphics[scale=0.47]{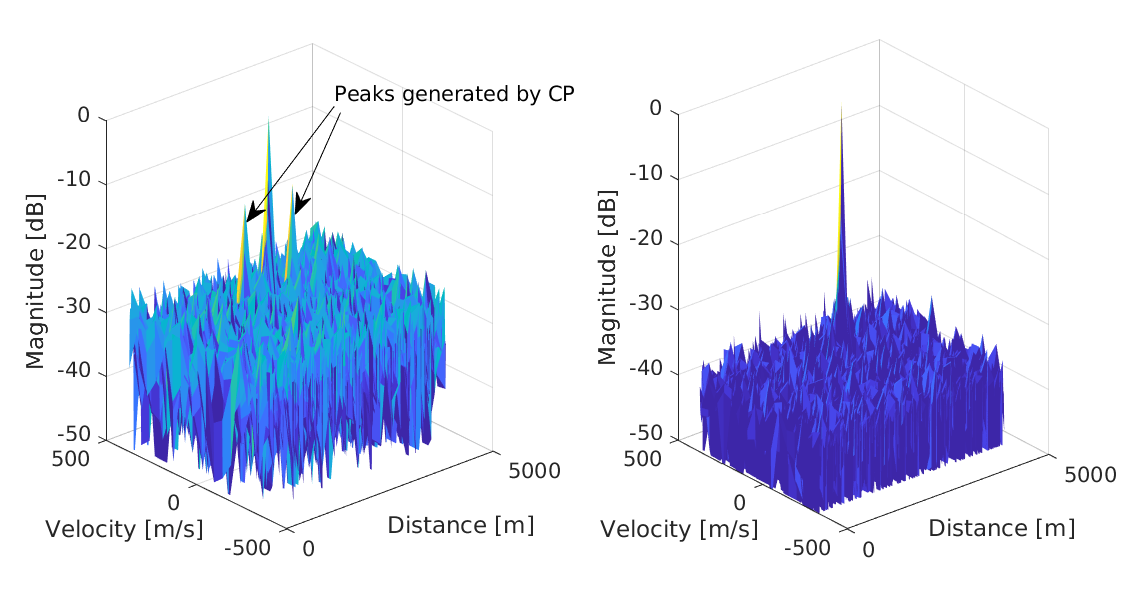}
\caption{DD map for correlation-based radar (left) and pilot based radar (right)}
\label{fig_radar_caf_pilot}
\end{figure}

Delay-Doppler (distance-velocity) radar maps for a target moving with velocity about 139~m/s (500~km/h), calculated for both tested radar approaches (correlation based and pilot based ones), are shown in figure~\ref{fig_radar_caf_pilot}. In both methods integration/observation time 100 milliseconds was used. Input signal had signal-to-noise ratio (SNR) equal to 0 decibels. In both cases one can clearly see sharp peaks in the delay-Doppler (distance-velocity) matrix which correspond to parameters of moving vehicles. However for CAF two additional lower peaks are visible which are generated by the CP of the pilot part of the RP-OTFS waveform. As in case of the pilot-based approach we eliminate CP from signal processing chain, such peaks are missing in DD map of this method. In case of correlation-based radar mean level of background side-lobes, surrounding the detection peak, is equal to about -30 decibels while for pilot-based radar -40 decibels.    

\begin{figure}[H]
\centering
\includegraphics[scale=0.47]{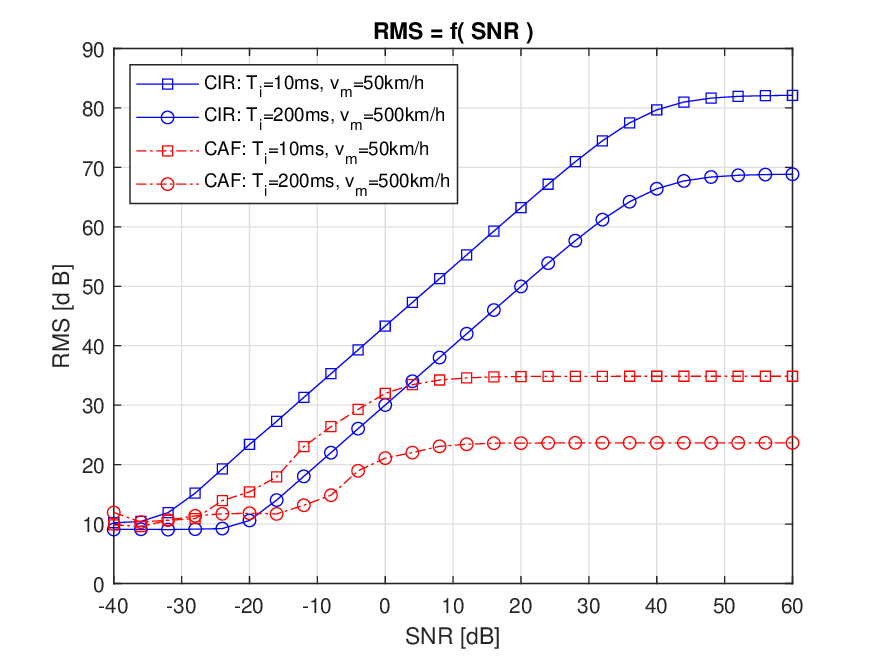}
\caption{Comparison of SNR curves for correlation and pilot-based radars}
\label{fig_inout_caf_cir}
\end{figure}

In figures~\ref{fig_inout_caf} and \ref{fig_inout_pilot} processing gain charts for both discussed RP-OTFS-based radars are shown, i.e. expressed in decibels root mean square (RMS) value of the method noise floor (visible in figure \ref{fig_radar_caf_pilot}) as a function of signal to noise ratio (SNR) of an input signal. Simulated maximum vehicle speed ($v_m$) was equal to 50 (13.9~m/s) albeit 500~km/h (139~m/s) and integration/observation time ($T_i$) was varying from 10~ms to 200~ms. In figure~\ref{fig_inout_caf_cir} both tested RP-OTFS-based radars are compared: it is seen that the pilot-based version outperforms the correlation-based one in DD detection hight, i.e. in noise robustness.  



\section{Discussion}

\begin{figure}[H]
\centering
\includegraphics[scale=0.47]{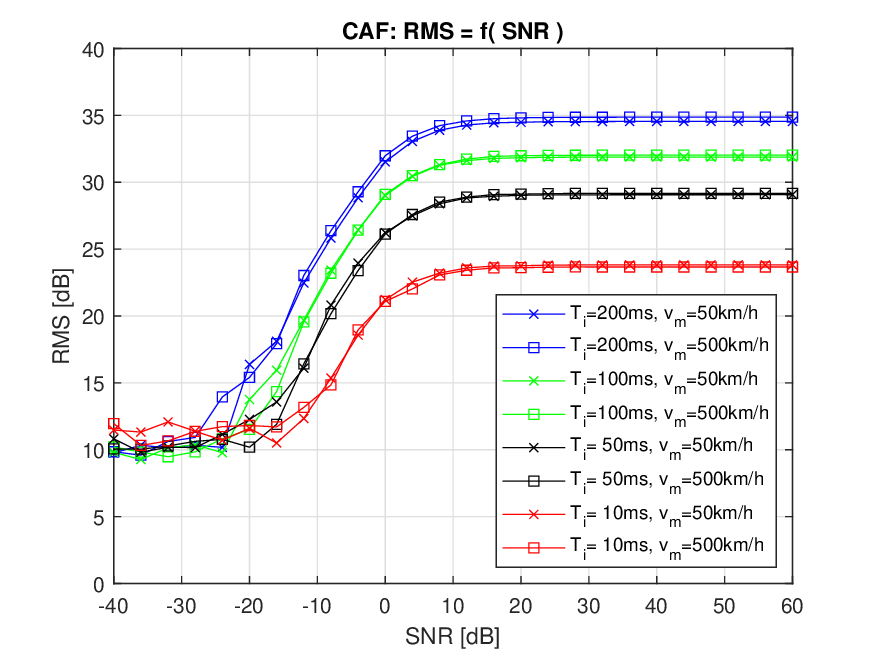}
\caption{SNR curves for correlation-based radar}
\label{fig_inout_caf}
\end{figure}

\begin{figure}[H]
\centering
\includegraphics[scale=0.47]{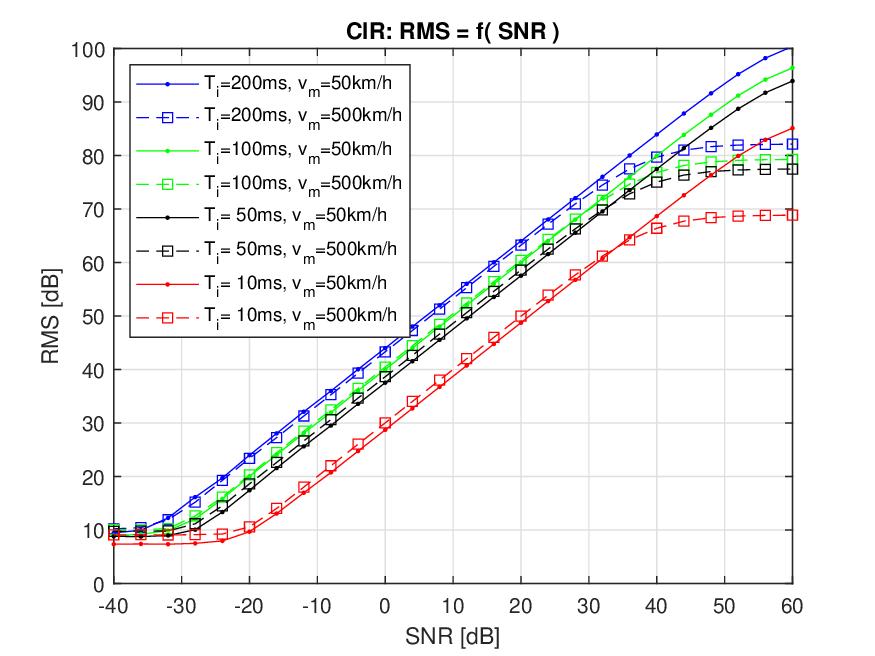}
\caption{SNR curves for pilot-based radar}
\label{fig_inout_pilot}
\end{figure}

The main limitation factor in case of the correlation based RP-OTFS radar is its high level of CAF side-lobes, resulting in significantly lower output SNR in comparison to the pilot-based radar. Figures ~\ref{fig_inout_caf} and \ref{fig_inout_pilot} confirm quantitative conclusions which can be drown from figure \ref{fig_radar_caf_pilot}.

As mentioned before, in the development of the discussed RP-OTFS-based ISAC system we have assumed that channel pulse response is quasi time-invariant in the pilot zone. In case of high-mobility Doppler channels this assumption is fulfilled only approximately. This fact will limit the maximum processing gain of the presented pilot based RP-OFTS radar. Consequences of this method drawback will increase for higher velocities as it is visible in fig.~\ref{fig_inout_pilot}. The same effect will be observed also when the pilot zone length will be increased. Nevertheless, obtained results confirm that the pilot-based radar outperforms the correlation-based one in terms of noise robustness.

\section{Conclusion}

Two moving vehicles detection approaches based on the RP-OTFS ISAC system were compared in this paper. The main limitation factor of the correlation based radar method is high level of CAF side-lobes, apart from existence of two additional peaks in CAF which are caused by repetition of the pilot samples. Detection of targets with low radar cross section on the background of strong background signal, so called clutter, e.g. direct path signal, is very challenging here. Presence of many ghost peaks in the delay-Doppler (distance-velocity) map makes subsequent processing steps in this method very challenging.
	
In turn, the pilot based  RP-OTFS radar is characterized by lower level of side-lobes in the delay-Doppler map and it does not have extra peaks caused by the repeating pilot samples. But this approach is sensitive to the quality of the channel impulse response estimation. In order to minimize error of the channel impulse response estimate, and in consequence error of the moving object detection, we need to keep pilot zone as short as possible.

\end{multicols}

\begin{thebibliography}{99}


\bibitem{6G_harsh}
H. Tataria et al., "6G Wireless Systems: Vision, Requirements, Challenges, Insights, Opportunities," Proc. IEEE, vol. 109, no. 7, pp. 1166-1199, July 2021.

\bibitem{6G_vision}
W. Saad, M. Bennis and M. Chen, "A Vision of 6G Wireless Systems: Applications, Trends, Technologies, and Open Research Problems," IEEE Network, vol. 34, no. 3, pp. 134-142, May/June 2020.

\bibitem{isac1}
F. Liu et al., ``Integrated Sensing and Communications: Toward Dual-Functional Wireless Networks for 6G and Beyond,'' IEEE J. on Selected Areas in Comm., vol. 40, no. 6, pp. 1728-1767, June 2022.

\bibitem{isac2}
Z. Wei et al., ``Integrated Sensing and Communication Signals Towards 5G-A and 6G: A Survey,'' IEEE Internet of Things Journal, early access, 2023.

\bibitem {ofdm_numerology}
Josue Flores de Valgas, Jose F. Monserrat, Hüseyin Arslan, "Flexible Numerology in 5G NR: Interference Quantification and Proper Selection Depending on the Scenario", Mobile Information Systems, vol. 2021, Article ID 6651326, 9 pages, 2021.

\bibitem {otfs1}
  R. Hadani et al., "Orthogonal Time Frequency Space Modulation," 2017 IEEE Wireless Comm. and Networking Conf. (WCNC), San Francisco, CA, USA, 2017, pp. 1-6, 2017.

\bibitem {otfs2}
Z. Wei at al., ``Orthogonal Time-Frequency Space Modulation: A Promising Next-Generation Waveform,'' IEEE Wireless Comm., vol. 28, iss. 4,
pp. 136-144, 2021.

\bibitem {my_rp1}
P. Karpovich and T. P. Zielinski, "Random-Padded OTFS Modulation for Joint Communication and Radar/Sensing Systems," 2022 23rd Int. Radar Symp. (IRS), pp. 104-109, Gdansk 2022.

\bibitem {my_rp2}
P. Karpovich et al., ``Field Tests of a Random-Padded OTFSM Waveform in a Joint Sensing and Communication System,'' IEEE ICC Int. Communications Conf., Rome 2023. 
  
\bibitem {zak}
  H. Bolcskei and F. Hlawatsch, "Discrete Zak transforms, polyphase transforms, and applications," in IEEE Trans. on Signal Processing, vol. 45, no. 4, pp. 851-866, April 1997.

\bibitem{radar}
M.A. Richards, “Fundamentals of Radar Signal Processing,” McGraw-Hill Education, 2014.

\bibitem {my_dvbt2}
 P. Karpovich et al., "Practical Results of Drone Detection by Passive Coherent DVB-T2 Radar," 21st Int. Radar Symp. (IRS), pp. 77-81, Warsaw 2020.
  
\bibitem {ofdm_base_radar}
M. Braun et al., "Parametrization of joint OFDM-based radar and comm. systems for vehicular applications," 2009 IEEE Int. Symp. on Personal, Indoor \& Mobile Radio Comm., pp. 3020-3024, Tokyo 2009.
  
\end{thebibliography}
\end{document}